\documentclass[10pt, lettersize,journal]{IEEEtran}
\usepackage{amsmath,amsfonts}
\usepackage{algorithmic}
\usepackage{array}
\usepackage[caption=false,font=normalsize,labelfont=sf,textfont=sf]{subfig}
\usepackage{textcomp}
\usepackage{stfloats}
\usepackage{url}
\usepackage{verbatim}
\usepackage{graphicx}
\def\BibTeX{{\rm B\kern-.05em{\sc i\kern-.025em b}\kern-.08em
		T\kern-.1667em\lower.7ex\hbox{E}\kern-.125emX}}
\usepackage{balance}
\begin{document}
	\newtheorem{theorem}{Theorem}
\newtheorem{definition}{Definition}
\newtheorem{lemma}{Lemma}
\newtheorem{example}{Example}
\newtheorem{proposition}{Proposition}
\newtheorem{corollary}{Corollary}
\newtheorem{remark}{Remark}
%\renewcommand\refname{References}	
%\renewcommand{\theequation}{\thesection.\arabic{equation}}
%\newcommand{\upcite}[1]{\textsuperscript{\textsuperscript{\cite{#1}}}}
%\catcode`@=11 \@addtoreset{equation}{section} \catcode`@=12
\title{The  equivalence of GRS codes and EGRS codes}

\author{ Canze Zhu
\thanks{The author is with the College of Mathematical Science, Sichuan Normal University, Chengdu Sichuan, 610066 (canzezhu@163.com).}}
        % <-this % stops a space}
% The paper headers
%\markboth{Journal of \LaTeX\ Class Files,~Vol.~14, No.~8, August~2021}%
%{Shell \MakeLowercase{\textit{et al.}}: A Sample Article Using IEEEtran.cls for IEEE Journals}

%\IEEEpubid{0000--0000/00\$00.00~\copyright~2021 IEEE}
% Remember, if you use this you must call \IEEEpubidadjcol in the second
% column for its text to clear the IEEEpubid mark.

	\maketitle
\begin{abstract}
 Generalized Reed-Solomon and extended generalized Reed-Solomon (abbreviation to GRS and EGRS) codes are the most well-known family of MDS codes with wide applications  in coding theory and practice.  Let $\mathbb{F}_q$ be the $q$ elements finite field, where $q$ is the power of a prime. For a linear code $\mathcal{C}$ over $\mathbb{F}_q$ with length $2\le n\le q$, we prove that $\mathcal{C}$ is a GRS code if and only if $\mathcal{C}$ is a EGRS code.
 
\end{abstract}

\begin{IEEEkeywords}Generalized Reed-Solomon codes;  Extended  generalized Reed-Solomon codes; Equivalence  
\end{IEEEkeywords}
\section{Introduction}
\IEEEPARstart{L}{et} $\mathbb{F}_q$ be the $q$ elements finite field, where $q$ is the power of a prime. An $[n,k,d]$ linear code $\mathcal{C}$ over $\mathbb{F}_q$ is a $k$-dimensional subspace of $\mathbb{F}_q^n$ with minimum (Hamming) distance $d$ and length $n$. If the
parameters reach the Singleton bound, namely, $d=n-k+1$, then $\mathcal{C}$  is  maximum distance separable (in short, MDS). MDS codes have various interesting properties and wide applications both in theoretical and practice, such as, combinatorial designs, finite geometry \cite{J1}, data storage and coding for distributed storage systems \cite{B2}. 
The most two well-known family of MDS codes are that of GRS and EGRS codes,  which not only have nice theoretic properties, but also have been extensively applied in engineering due to their easy encoding and efficient decoding algorithm. Similarly, self-dual or small hull codes are another interesting class of linear codes due to their nice algebraic structure, their wide applications in cryptography (in particular secret sharing) \cite{C3} and constructions of quantum MDS codes \cite{C4,M5}.

Therefore, the constructions of the self-dual, or small hull MDS codes  have attracted a lot of attention in recent years. There are a lot of work for these MDS codes via generalized Reed-Solomon (GRS) codes and their extended codes\cite{6,7,8,9,10,11,12,13}. It is worth mentioning that the authors also distinguish the GRS code and the EGRS code, and presented the similar method to construct the self-dual or small hull GRS code and EGRS code over $\mathbb{F}_q$ with length less than or equal to $q$. Especially, Fang et. al prove that all the self-dual EGRS codes over $\mathbb{F}_q$ with length less than or equal to $q$ can be constructed from GRS codes with the same parameters \cite{13}.

In this letter, for a linear code $\mathcal{C}$ over $\mathbb{F}_q$ with length $2\le n\le q$,  we prove that $\mathcal{C}$ is a GRS code if and only if $\mathcal{C}$ is a EGRS code. This result implies that there is unnecessary to distinguish the GRS code and the EGRS code when the length less than or equal to $q$.
\section{ Main Result}
In the following, we  review the definitions of the GRS code and the EGRS code, respectively.

\begin{definition}[\cite{J1}, Ch. $10$. $\S8$ \& Ch. $11$. $\S5$]\label{d0} 
	For $k,n\in\mathbb{N}$ with $1\le k\le n\le q$, let $\boldsymbol{v} = (v_1,\ldots,v_{n})\in (\mathbb{F}_q^{*})^{n}$, $\boldsymbol{\alpha}=(\alpha_1,\ldots,\alpha_n)\in\mathbb{F}_q^n$ with $\alpha_i\neq \alpha_j$ $(i\neq j)$ and $$\mathcal{V}_{k}=\{f(x)~|~f(x)\in\mathbb{F}_q[x]\text{ ~and ~}\deg f(x)\le k-1\}.$$ Then
	the GRS code is defined as
	\begin{align*}
	\mathcal{GRS}_{k,n}(\boldsymbol{\alpha},\boldsymbol{v})&=\{(v_1f({\alpha_{1}}),\ldots,v_nf(\alpha_{n}))\big|f(x)\in\mathcal{V}_k\}.
	\end{align*}	
	The EGRS code is defined as
	\begin{align*}
	\mathcal{EGRS}_{k,n}(\boldsymbol{\alpha},\boldsymbol{v})\!=\!\{(v_1f({\alpha_{1}}),\ldots,v_nf(\alpha_{n}),f_{k-1})\big|f(x)\!\in\!\mathcal{V}_k\},
	\end{align*}	
	where $f_{k-1}$ is a coefficient of $x^{k-1}$ in $f(x)$.\\
\end{definition}

Before giving the main result, we fix the following two nonations.
\begin{align*}
\mathrm{GRS}\!=\!\big\{ \mathcal{C}\!~|\!~\mathcal{C}~\text{is a GRS code over $\mathbb{F}_q$ with length~} n\ge 2\big\}.
\end{align*}
\begin{align*}
\mathrm{EGRS}\!=\!\big\{ \mathcal{C}\!~|\!~\mathcal{C}~\text{is a EGRS code over $\mathbb{F}_q$ with length~} n\le q \big\}.
\end{align*}

The following theorem presents the main result of this paper.\\

\begin{theorem}\label{t1} $\mathrm{GRS}=	\mathrm{EGRS}.$\\
\end{theorem}

The proof of Theorem \ref{t1} is given in the next section.\\

\begin{remark} It follows from Theorem $1$ that the GRS code and the EGRS code are equivalent to each other when length $2\le n\le q$, which implies that many known results of GRS codes are aways suitable for EGRS codes directly.
\end{remark}

\section{Proofs}
 The following lemma that will be used to prove that $	\mathrm{GRS}\subseteq\mathrm{EGRS}$.
\begin{lemma}\label{L1}
	For $k,n\in\mathbb{N}$ with $1\le k\le n\le q$, let  $a\in\mathbb{F}_q$, $\lambda\in\mathbb{F}_q^{*}$,  $\boldsymbol{\alpha}=(\alpha_1,\ldots,\alpha_n)\in\mathbb{F}_q^n$ with $\alpha_i\neq \alpha_j$ $(i\neq j)$ and  $\boldsymbol{v} = (v_1,\ldots,v_{n})\in (\mathbb{F}_q^{*})^{n}$. If $\boldsymbol{\beta}=(\alpha_1-a,\ldots,\alpha_n-a)$ and $\boldsymbol{w} = (\lambda v_1,\ldots,\lambda v_{n})$, then
	\begin{align*}
	\mathcal{GRS}_{k,n}(\boldsymbol{\alpha},\boldsymbol{v})=\mathcal{GRS}_{k,n}(\boldsymbol{\beta},\boldsymbol{w}).
	\end{align*}
\end{lemma}

{\bf \emph{Proof}}. For any $\big(v_1f(\alpha_1),\ldots,v_nf(\alpha_n)\big)\in\mathcal{GRS}_{k,n}(\boldsymbol{\alpha},\boldsymbol{v})$ with $f(x)=\sum\limits_{i=0}^{k-1}f_i x^{i}\in\mathbb{F}_q[x]$, let $g(x)=\lambda^{-1} f(x+a)$, then $g(x)\in\mathbb{F}_q[x]$ with  $\deg g(x)\le k-1$. And then, 
for any $i\in\{1,\ldots,n\}$, we have
\begin{align*}
v_i f(\alpha_i)&=\lambda v_i \big(\lambda^{-1}f\big((\alpha_i-a)+a\big)\big)=\lambda v_i g(\alpha_i-a),
\end{align*}
that is
\begin{align*}
\big(v_1f(\alpha_1),\ldots,v_nf(\alpha_n)\big)=\big(\lambda v_1 g(\alpha_1\!-a),\ldots,\lambda v_n g(\alpha_n\!-a)\big),
\end{align*}
which leads 
\begin{align*}
\mathcal{GRS}_{k,n}(\boldsymbol{\alpha},\boldsymbol{v})\subseteq\mathcal{GRS}_{k,n}(\boldsymbol{\beta},\boldsymbol{w}).
\end{align*}
Note that \begin{align*}
\dim(
\mathcal{GRS}_{k,n}(\boldsymbol{\alpha},\boldsymbol{v}))=\dim(\mathcal{GRS}_{k,n}(\boldsymbol{\beta},\boldsymbol{w})),
\end{align*} thus
\begin{align*}
\mathcal{GRS}_{k,n}(\boldsymbol{\alpha},\boldsymbol{v})=\mathcal{GRS}_{k,n}(\boldsymbol{\beta},\boldsymbol{w}).
\end{align*}
$\hfill\Box$
\begin{remark}\label{REM1}
By taking $a=\alpha_n$ and $\lambda=v_n^{-1}$ in Lemma \ref{L1}, then $\alpha_n-a=0$ and $\lambda v_n=1$, thus for any GRS code $\mathcal{GRS}_{k,n}(\boldsymbol{\alpha},\boldsymbol{v})$, we can alway assume
	$\alpha_n=0$ and $v_n=1$ without loss of generality.\\
\end{remark}

Now, $\mathrm{GRS}\subseteq \mathrm{EGRS}$ and 
$\mathrm{EGRS}\subseteq \mathrm{GRS}$ are given in Lemmas $\ref{l2}$-$\ref{l3}$, respectively.\\

\begin{lemma}\label{l2}
		$\mathrm{GRS}\subseteq \mathrm{EGRS}.$	
\end{lemma}

	{\bf \emph{Proof}}. By Remark \ref{REM1}, for any $\mathcal{C}\in \mathrm{GRS}$, there exists a $2\le n\le q$, $\boldsymbol{v}=(v_1,\ldots,v_{n-1},1)\in (\mathbb{F}_q^{*})^{n}$ and $\boldsymbol{\alpha}=(\alpha_1,\ldots,\alpha_{n-1},0)\in\mathbb{F}_q^n~\text{with}~$ $\alpha_i\neq0,~ \alpha_i\neq \alpha_j~(i\neq j)$ such that $\mathcal{C}=\mathcal{GRS}_{k,n}(\boldsymbol{\alpha},\boldsymbol{v})$. Now we find some $\boldsymbol{\beta},\boldsymbol{w}\in\mathbb{F}_q^{n-1}$ such that
	\begin{align}\label{R2}
	\mathcal{GRS}_{k,n}(\boldsymbol{\alpha},\boldsymbol{v})=\mathcal{EGRS}_{k,n-1}(\boldsymbol{\beta},\boldsymbol{w}).
	\end{align}

	By $\alpha_i\neq 0$ $(i=1,\ldots,n-1)$, we can choose
	$$\boldsymbol{\beta}=(\beta_1,\ldots,\beta_{n-1})=\big(\alpha_1^{-1},\ldots,\alpha_{n-1}^{-1}\big)$$ and
	$$\boldsymbol{w}=(w_1,\ldots,w_{n-1})=\big(v_1\alpha_1^{k-1},\ldots,v_{n-1}\alpha_{n-1}^{k-1}\big).$$
	It is easy to see that $\beta_i\neq \beta_{j}$ $(i\neq j)$ and $\boldsymbol{w}\in(\mathbb{F}_q^{*})^{n-1}$. Now we show that $(\ref{R2})$ holds.
	
	For any $\!(v_1f(\alpha_1),\ldots, v_{n-1}f(\alpha_{n-1}),f(0))\!\in\!\mathcal{GRS}_{k,n}(\!\boldsymbol{\alpha},\!\boldsymbol{v})$
	 with $f(x)=\sum\limits_{i=0}^{k-1}f_{i}x^{i}\in\mathbb{F}_q[x]$, let $$g(x)=\sum\limits_{i=0}^{k-1}f_{i}x^{k-1-i}.$$ It is easy to see that $g(x)\in\mathbb{F}_q[x]$ with $\deg g(x)\le k-1.$ 
	Furthermore, the coefficient of $x^{k-1}$ in $g(x)$ is $$g_{k-1}=f_0=f(0),$$ and for any $j\in\{1,\ldots,n-1\}$
	\begin{align*}
	w_{j}g(\beta_{j})=v_j\alpha_j^{k-1}\sum\limits_{i=0}^{k-1}f_{i}\alpha_j^{i-(k-1)}=v_jf(\alpha_j).
	\end{align*}
	Thus
	\begin{align*}
	&\big(v_1f(\alpha_1),\ldots,v_{n-1}f(\alpha_{n-1})f(0)\big)\\
	=&\big(w_1 g(\beta_1),\ldots,w_{n-1}g(\beta_{n-1}),g_{k-1}\big),
	\end{align*}
	which leads\begin{align*}
	\mathcal{GRS}_{k,n}(\boldsymbol{\alpha},\boldsymbol{v})\subseteq\mathcal{EGRS}_{k,n-1}(\boldsymbol{\beta},\boldsymbol{w}).
	\end{align*}Note that $$\dim\big(\mathcal{GRS}_{k,n}(\boldsymbol{\alpha},\boldsymbol{v})\big)=\dim\big(\mathcal{EGRS}_{k,n-1}(\boldsymbol{\beta},\boldsymbol{w})\big),$$ thus $(\ref{R2})$ holds.

	$\hfill\Box$
	
\begin{lemma}\label{l3}
	$\mathrm{EGRS}\subseteq \mathrm{GRS}.$	
\end{lemma}

{\bf \emph{Proof}}. 
For any $\mathcal{C}\in \mathrm{EGRS}$, there exists a $n\!\le\! q-1$, $\!\boldsymbol{v}\!=\!(v_1,\ldots,v_n)\in (\mathbb{F}_q^{*})^{n}\!$ and $\boldsymbol{\alpha}\!=\!(\alpha_1,\ldots,\alpha_n)\!\in\!\mathbb{F}_q^n$ with $\alpha_i\neq \alpha_j~(i\neq j)$ such that $\mathcal{C}=\mathcal{EGRS}_{k,n}(\boldsymbol{\alpha},\boldsymbol{v}) $. Now we find a $\boldsymbol{\beta},\boldsymbol{w}\in\mathbb{F}_q^{n+1}$ such that
\begin{align}\label{R1}
\mathcal{GRS}_{k,n+1}(\boldsymbol{\beta},\boldsymbol{w})=\mathcal{EGRS}_{k,n}(\boldsymbol{\alpha},\boldsymbol{v}).
\end{align}

By $n\le q-1$, we know that $A=\mathbb{F}_{q}\backslash \{\alpha_1,\ldots,\alpha_n\}$ is not empty, by choosing $\gamma\in A$ arbitrarily, let
\begin{align*}
	\boldsymbol{\beta}&=(\beta_1,\ldots,\beta_n,\beta_{n+1})\\
	&=\big((\alpha_1-\gamma)^{-1},\ldots,(\alpha_n-\gamma)^{-1},0\big)
\end{align*}and
\begin{align*}
\boldsymbol{w}=&(w_1,\ldots,w_n,w_{n+1})\\
=&\big(v_1(\alpha_1-\gamma)^{k-1},\ldots,v_n(\alpha_n-\gamma)^{k-1},1\big).
\end{align*} 
It is easy to see that $\beta_i\neq \beta_{j}$ $(i\neq j)$ and $\boldsymbol{w}\in(\mathbb{F}_q^{*})^{n+1}$. Now we show that $(\ref{R1})$ holds.

For any $(v_1f(\alpha_1),\ldots,v_nf(\alpha_n),f_{k-1})\in	\mathcal{EGRS}_{k,n}(\boldsymbol{\alpha},\boldsymbol{v})$ with $f(x)=\sum\limits_{i=0}^{k-1}f_{i}x^{i}\in\mathbb{F}_q[x]$, let $$g(x)=\sum\limits_{i=0}^{k-1}f_{i}(1+\gamma x)^ix^{k-1-i}.$$ It is easy to see that $g(x)\in\mathbb{F}_q[x]$ with $\deg g(x)\le k-1.$ 
Furthermore, 
\begin{align*}
w_{n+1}g(\beta_{n+1})=g(0)=f_{k-1},
\end{align*}and for any $j\in\{1,\ldots,n\}$, we have
\begin{align*}
&w_jg(\beta_j)\\
=&v_j(\alpha_j-\gamma)^{k-1}\sum_{i=0}^{k-1}f_{i}(1+\gamma (\alpha_j-\gamma)^{-1})^{i}(\alpha_j-\gamma)^{-(k-1-i)}\\
=&v_j(\alpha_j-\gamma)^{k-1}\sum_{i=0}^{k-1}f_{i}(\alpha_j(\alpha_j-\gamma)^{-1})^{i}(\alpha_j-\gamma)^{-(k-1-i)}\\
=&v_jf(\alpha_j),
\end{align*}
that is
\begin{align*}
&\big(w_1g(\beta_1),\ldots,w_ng(\beta_{n}),w_{n+1}g(\beta_{n+1})\big)\\
=&\big(v_1f(\alpha_1),\ldots,v_nf(\alpha_n),f_{k-1}\big),
\end{align*}
which leads\begin{align*}
\mathcal{EGRS}_{k,n}(\boldsymbol{\alpha},\boldsymbol{v})\subseteq\mathcal{GRS}_{k,n+1}(\boldsymbol{\beta},\boldsymbol{w}).
\end{align*}Note that $$\dim\big(\mathcal{EGRS}_{k,n}(\boldsymbol{\alpha},\boldsymbol{v})\big)=\dim\big(\mathcal{GRS}_{k,n+1}(\boldsymbol{\beta},\boldsymbol{w})\big),$$ thus $(\ref{R1})$ holds.

 $\hfill\Box$
 
{\bf \emph{Proof of Theorem $\ref{t1}$}.} The Theorem $\ref{t1}$ follows from Lemmas $\ref{l2}$-$\ref{l3}$ directly.

$\hfill\Box$
\section{Conclusion}
In this letter, for a linear code $\mathcal{C}$ with length $2\le n\le q$, we prove that $\mathcal{C}$ is a GRS code if and only if $\mathcal{C}$ is a EGRS code.


\begin{thebibliography}{1}
%\bibitem{B1} Huffman W., Pless V., Fundamentals of Error Correcting Codes, Cambridge University Press, Cambridge (2003).
\bibitem{B2} Balaji, S. Krishnan, M.  Vajha, M. Ramkumar, V.  Sasidharan, B.  and  Kumar, P. Erasure coding for distributed storage: An overview. Science China Information Sciences 61.10 (2018): 1-45.

\bibitem{12}Cao, Meng.  MDS Codes With Galois Hulls of Arbitrary Dimensions and the Related Entanglement-Assisted Quantum Error Correction.  IEEE Transactions on Information Theory 67.12 (2021): 7964-7984.
\bibitem{C4} Calderbank, A. Robert, and Peter W. Shor.  Good quantum error-correcting codes exist.  Physical Review A 54.2 (1996): 1098.
\bibitem{C3} Cramer, R. Daza, V.  Gracia, I.  Urroz, J. Leander, G. Martí-Farré, J. Padró, C.   On codes, matroids, and secure multiparty computation from linear secret-sharing schemes.  IEEE Transactions on Information Theory 54.6 (2008): 2644-2657.

\bibitem{6} Fang, Weijun, Xia, Shutao and Fu, Fangwei.  Construction of MDS Euclidean self-dual codes via two subsets.  IEEE Transactions on Information Theory 67.8 (2021): 5005-5015.
\bibitem{13} Fang, Weijun, Xia, Shutao and Fu, Fangwei.   A note on self-dual generalized Reed-Solomon codes.  arXiv preprint arXiv:2005.11732 (2020).
\bibitem{7} Fang, Xiaolei,  Liu, Meiqing and Luo, Jinquan.  New MDS Euclidean self-orthogonal codes.  IEEE Transactions on Information Theory 67.1 (2020): 130-137.

\bibitem{11} Fang, Xiaolei, Jin, Renjie, and Luo, Jinquan.  New Galois Hulls of GRS Codes and Application to EAQECCs.  Cryptography and Communications 14.1 (2022): 145-159.
\bibitem{9} Guo, Guanmin, and Li, Ruihu.  Hermitian Self-Dual GRS and Extended GRS Codes.  IEEE Communications Letters 25.4 (2020): 1062-1065.
\bibitem{8} Huang, Ziteng,  Fang, Weijun and Fu, FangWei.  New constructions of MDS self-dual and self-orthogonal codes via GRS codes.  arXiv preprint arXiv:2103.11665 (2021).
\bibitem{10} Ning, Yu, Ye, Zuo, Ge, Gennian, Miao, Fuyou, and  Zhang Xiande.  New Results on Self-Dual Generalized Reed-Solomon Codes.  IEEE Transactions on Information Theory 67.11 (2021): 7240-7252.
\bibitem{J1} MacWilliams, Florence Jessie, and Neil James Alexander Sloane. The theory of error correcting codes. Vol. 16. Elsevier, 1977.
\bibitem{M5} Steane, Andrew M.  Simple quantum error-correcting codes.  Physical Review A 54.6 (1996): 4741.

\end{thebibliography}
\end{document}